\documentclass[a4paper,fleqn]{cas-sc}

\usepackage[numbers]{natbib}

\usepackage{xcolor}

\begin{document}
\let\WriteBookmarks\relax
\def\floatpagepagefraction{1}
\def\textpagefraction{.001}

\noindent\textit{Accepted manuscript of an article
published in Optics \& Laser Technology.}

\noindent The final published version is available at
\url{https://doi.org/10.1016/j.optlastec.2026.116662}.

\noindent 
This manuscript version is made available under
the CC BY-NC-ND 4.0 license:
\url{https://creativecommons.org/licenses/by-nc-nd/4.0/}.
\shorttitle{Compact deterministic liquid-crystal polarization controller}

\shortauthors{Ren-De Liu et~al.}

\title [mode = title]{Compact deterministic liquid-crystal polarization controller with single-measurement direct control}                      




\author[1,2]{Ren-De Liu}\fnmark[1]
\author[1]{Jie Wang}\fnmark[1]
\author[1]{Dong-Dong Li}
\author[1]{Xian-Chang Du}
\author[1]{Dao-Zhong Xiao}
\author[1]{Shuai Li}
\author[1]{Chuan-Yang Ding}
\author[1]{Yan-Lin Tang}
\author[2,3,4]{Yang Li}
\author[2,3,4]{Sheng-Kai Liao}
\author[2,4]{Teng-Yun Chen}
\author[1]{Shi-Biao Tang} \cormark[1]

\affiliation[1]{organization={QuantumCTek Co., Ltd.},
    city={Hefei},
    postcode={230088}, 
    country={China}}

\affiliation[2]{organization={Hefei National Research Center for Physical Sciences at the Microscale and School of Physical Sciences, University of Science and Technology of China},
        city={Hefei},
    postcode={230026}, 
    country={China}}

\affiliation[3]{organization={Shanghai Research Center for Quantum Science and CAS Center for Excellence in Quantum Information and Quantum Physics, University of Science and Technology of China},
        city={Shanghai},
    postcode={201315}, 
    country={China}}

\affiliation[4]{organization={Hefei National Laboratory, University of Science and Technology of China},
        city={Hefei},
    postcode={230088}, 
    country={China}}

\cortext[cor1]{Corresponding author: shibiao.tang@quantum-info.com}
\fntext[1]{These authors contributed equally to this work.}

\begin{abstract}
Polarization controllers (PCs) and their control algorithms are critical for compensating environmental disturbances and maintaining signal fidelity in polarization-sensitive photonic systems. However, existing controllers often involve trade-offs among key performance metrics, while associated control algorithms either converge inefficiently or require multiple state-of-polarization (SOP) measurements. This work presents a compact deterministic liquid-crystal polarization controller (LC-PC) with balanced overall performance, featuring low insertion loss (<1 dB), low driving voltage (<6 V), and full Poincaré sphere coverage. The device exhibits highly predictable and repeatable polarization modulation characterized by SOP-independent trajectory normal vectors,  enabling a single-measurement direct-calculation algorithm using pre-characterized rotation axes and voltage-to-rotation relationships, with degree-level polarization-control accuracy. The combination of this compact deterministic LC-PC and its efficient control scheme provides a practical solution for polarization management in quantum key distribution, optical sensing, and advanced photonic systems.

\end{abstract}




\maketitle

\section{Introduction}
State-of-polarization (SOP) fluctuations induced by environmental perturbations can severely degrade the performance of polarization-sensitive photonic systems, including quantum communication \cite{yinEntanglementbasedSecureQuantum2020,liuContinuousvariableQuantumKey2020a,yinPolarizationCompensationEntanglementbased2025a,peranicStudyPolarizationCompensation2023,wuResearchRealtimePolarization2022,liHighrateQuantumKey2023a,duSiliconbasedDecoderPolarizationencoding2023,weiResourceefficientQuantumKey2023a}, optical sensing \cite{chenEnhancingWeakDisturbance2025,tangDistributedOpticalFiber2024}, and biomedical imaging \cite{hePolarisationOpticsBiomedical2021a,guanReviewPolarizationbasedTechnology2025}. Polarization controllers (PCs) therefore play a crucial role in dynamically manipulating and stabilizing SOPs to maintain system performance. The practical deployment of PCs depends on multiple device-level metrics, including insertion loss, response speed, driving voltage, footprint, and the predictability of the output SOP.

Existing PC technologies exhibit inherent trade-offs among these performance metrics. Waveplate-based PCs provide accurate polarization control but suffer from bulky structures and slow response due to mechanical rotation \cite{shao-kaiRealizationArbitraryInverse2007,tanPolarizationCompensationMethod2023,goldsteinAnalysisPolarizedLight1970a,luoResearchPolarizationCompensation2024}. Fiber-squeezer PCs are compact and fast but are affected by hysteresis and creep, which degrade modulation repeatability \cite{aartsNewEndlessPolarization1989,hirabayashiFeedforwardContinuousComplete2003,huAdaptivePolarizationControl2023}. Chip-based PCs offer extremely small footprints and high-speed polarization-control capability but may incur additional fiber-to-chip coupling loss\cite{sarmiento-merenguelDemonstrationIntegratedPolarization2015,kimIntegratedopticPolarizationControllers2012,shahwarPolarizationManagementSilicon2024a,linHighperformancePolarizationManagement2022a, gaoUltrafastResetfreeIntegrated2026}. Liquid-crystal polarization controllers (LC-PCs) provide stable polarization modulation, low driving voltage, and low insertion loss, making them attractive candidates for practical polarization control applications \cite{zhuangPolarizationControllerUsing1999,jimenez-girelaPolarizationcorrectionDeviceLiquid2025a,hanProgrammableLinearRetarder2025}. Nevertheless, many reported LC-PCs are still implemented using discrete bulk-optic assemblies, which limits their compactness and integration capability.

In addition to device performance, practical polarization control also relies on efficient control algorithms. Polarization control algorithms can generally be classified into gradient-based \cite{youngComparisonOptimisationAlgorithms2026,zhangParticleSwarmOptimization2005,maAutomatedControlAlgorithms2020} and direct-calculation-based \cite{tanRealtimePolarizationCompensation2024a} approaches. Gradient-based algorithms rely on iterative optimization and typically require tens or hundreds of iterations to converge. Direct-calculation approaches can achieve substantially faster convergence by directly solving for the required polarization transformation. However, the effectiveness of these approaches critically depends on the PC exhibiting a deterministic and predictable response, which enables an accurate mapping between the applied control signals and the resulting polarization states. Existing implementations typically establish the required polarization transformation through measurements of the transmission channel, such as reconstructing the channel Mueller matrix from multiple SOP measurements \cite{luoResearchPolarizationCompensation2024,jimenez-girelaPolarizationcorrectionDeviceLiquid2025a}. These repeated online measurements and channel-characterization procedures increase system complexity and initialization time, limiting the practical implementation of direct-calculation polarization control.

To address these challenges, we demonstrate a compact LC-PC in a butterfly package (30 $\times$ 12.7 $\times$ 13 mm$^3$). The device achieves low insertion loss (<1 dB), low driving voltage (<6 V), low polarization-dependent loss (PDL), and full Poincaré sphere coverage, while exhibiting highly deterministic polarization modulation with SOP-independent trajectory normal vectors. Leveraging these characteristics, a direct-calculation control algorithm is developed that uses pre-characterized device parameters and requires only a single SOP measurement for control, achieving degree-level polarization-control accuracy without iterative optimization. The proposed compact deterministic LC-PC, together with its efficient single-measurement control scheme, provides a practical solution for polarization management in loss- and space-constrained photonic systems.

\section{Device fabrication and characterization}
\subsection{Fabrication and implementation}

\begin{figure*}
	\centering
	\includegraphics[width=0.8\textwidth]{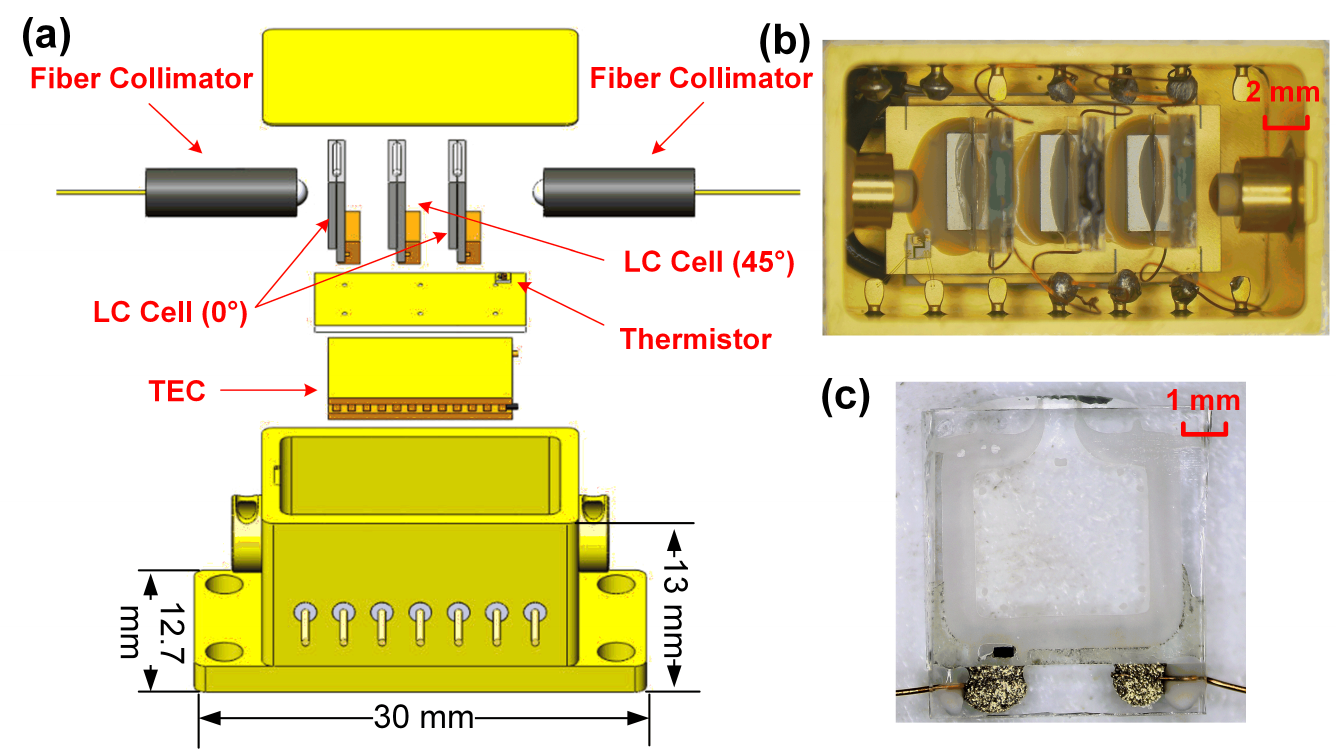}
	\caption{Structural diagram of LC-PC: (a) Exploded View of LC- PC Structure; (b) Internal layout of the LC-PC; (c) Liquid crystal cell.}
	\label{FIG:1}
\end{figure*}

Figure~\ref{FIG:1} illustrates the structure and packaging of the proposed LC-PC. As shown in Fig.~\ref{FIG:1}(a), the device is realized in a compact butterfly package with dimensions of $30\text{mm} \times 12.7\text{mm} \times 13\text{mm}$. The polarization modulation unit consists of three miniature LC cells with fast-axis orientations of $0^\circ$, $45^\circ$, and $0^\circ$, respectively. A photograph of the assembled device after removing the package cover is shown in Fig.~\ref{FIG:1}(b), and the structure of an individual LC cell is presented in Fig.~\ref{FIG:1}(c).
 
Each LC cell has dimensions of $6.5\text{mm} \times 5\text{mm} \times 1.1\text{mm}$ and consists of a nematic liquid-crystal layer sandwiched between two indium tin oxide (ITO)-coated glass substrates. An anti-parallel alignment configuration is formed through a polyimide rubbing process, while a uniform cell gap of $10\mu\text{m}$ is maintained using precision spacers. To reduce optical loss, anti-reflection coatings are applied to the outer surfaces of the LC cells. The electrodes of each cell are independently connected to the package pins, allowing individual voltage control of the three modulation channels. Optical coupling between the input and output fibers is achieved using a pair of fiber collimators. All optical components are actively aligned and fixed using UV- and thermally-cured epoxy to ensure stable optical coupling and long-term mechanical reliability.

Because the polarization response of the LC-PC is temperature dependent, integrated thermistors and thermoelectric coolers (TECs) are employed for active thermal stabilization to maintain a stable internal operating temperature against ambient-temperature variations. The package is subsequently hermetically sealed using parallel-seam welding technology, providing enhanced mechanical robustness and environmental reliability.

\subsection{Theoretical analysis and simulation}
Liquid crystals exhibit anisotropic dielectric and optical properties due to their elongated molecular structure, resulting in polarization-dependent propagation characteristics. When polarized light propagates through a liquid-crystal layer, it can be decomposed into ordinary and extraordinary components. The phase delay accumulated between these two components modifies the SOP of the transmitted light. By adjusting the amplitude of the applied electric field, the orientation of liquid-crystal molecules changes, leading to a voltage-dependent birefringence and consequently an electrically controllable phase retardation \cite{Yeh1999}.

The liquid-crystal cell can therefore be modeled as a voltage-controlled linear retarder. Based on Jones calculus, its polarization transformation matrix can be expressed as

\begin{equation}
M_{LC}(\delta, \theta) = 
\begin{bmatrix}
\cos^2 \theta + e^{-i\delta} \sin^2 \theta & \cos \theta \sin \theta (1 - e^{-i\delta}) \\
\cos \theta \sin \theta (1 - e^{-i\delta}) & \sin^2 \theta + e^{-i\delta} \cos^2 \theta
\end{bmatrix}
\end{equation}

\begin{equation}
\delta = \frac{2\pi}{\lambda} \Delta n(V) d
\end{equation}
where $\theta$ represents the orientation of the fast axis of the liquid crystal cell, $\delta$ is the phase retardation, $d$ is the liquid-crystal layer thickness, $\lambda$ is the optical wavelength, and $\Delta n(V)$ represents the voltage-dependent effective birefringence.

\begin{figure*}
	\centering
	\includegraphics[width=1.0\textwidth]{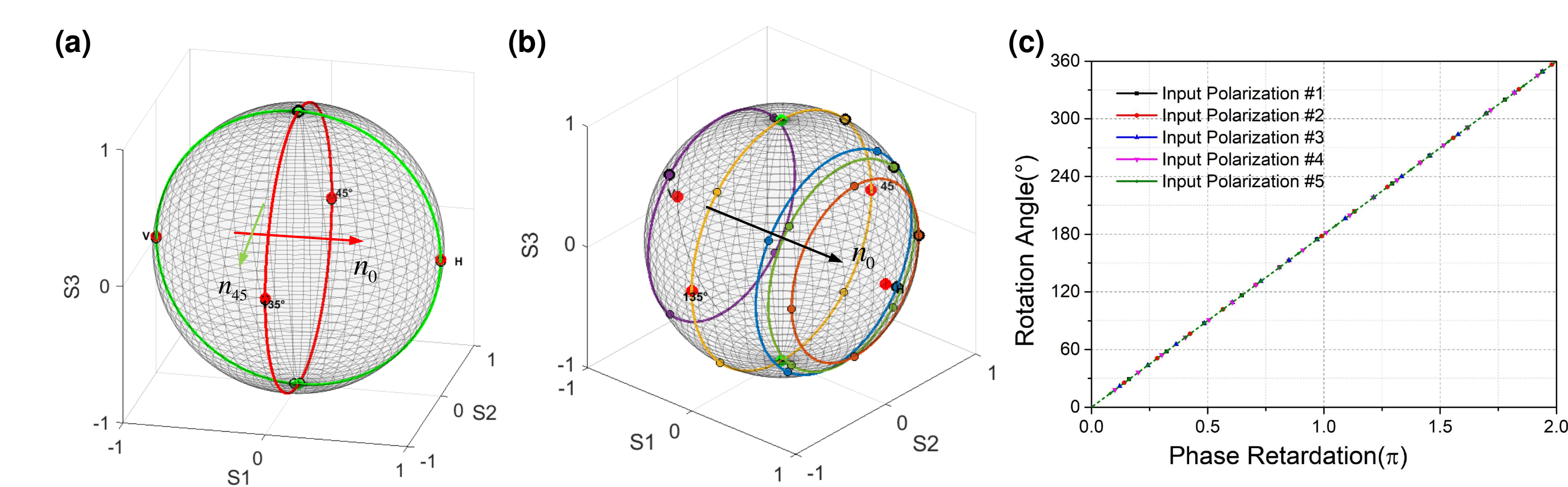}
	\caption{Polarization modulation characteristics: (a) Modulation trajectories for 0° and 45° fast-axis orientations on the Poincaré sphere; (b) Trajectory circles corresponding to different input SOPs; (c) Phase retardation vs. rotation angle for various input polarization states.}
	\label{FIG:2}
\end{figure*}

Simulations of the Jones matrices for liquid crystal cells with fast axes at $0^\circ$ and $45^\circ$ reveal their polarization modulation characteristics on the Poincaré sphere (Fig.~\ref{FIG:2}(a)). As the phase retardation varies, the SOP traces a circular trajectory. The trajectories generated by the two cells are mutually orthogonal, as are their normal vectors $\vec{n}_0$ and $\vec{n}_{45}$. Under a fixed liquid crystal cell fast-axis orientation, the modulation characteristics were analyzed for different input SOPs. As shown in Fig.~\ref{FIG:2}(b), each input SOP follows a distinct circular trajectory on the Poincaré sphere. Although these trajectories differ, they share a common normal vector $\vec{n}_0$, confirming that the vector is intrinsic to the liquid crystal cell and independent of the input SOP. Moreover, varying the phase retardation causes each SOP to rotate about its own center on the sphere. By taking the initial SOP as a reference point, a deterministic relationship between the phase retardation and the polarization rotation angle can be established, as illustrated in Fig.~\ref{FIG:2}(c). Importantly, the rotation angle is uniquely determined by the phase retardation and remains independent of the input SOP. This deterministic phase-retardation-to-rotation relationship forms the physical basis for the proposed single-measurement direct-calculation polarization control algorithm.

\subsection{Performance characterization}
At the nominal operating wavelength of 1550 nm, the fabricated LC-PC exhibits an insertion loss of $0.97\text{ dB}$ and a PDL of only $0.04\text{ dB}$. The driving voltages required to achieve a $2\pi$ phase retardation are $5.9\text{ V}$, $5.8\text{ V}$, and $5.95\text{ V}$ for CH1, CH2, and CH3, respectively. The measured response time is $102\text{ ms}$ for an approximately $2\pi$ phase-retardation transition at $40^\circ\text{C}$. The response becomes faster at higher driving voltages and elevated operating temperatures.

By varying the voltages applied to the three channels of the LC-PC, different phase retardation can be generated, thereby achieving polarization modulation. The polarization modulation characteristics and the deterministic response of the LC-PC were experimentally characterized by measuring the SOP evolution on the Poincaré sphere.

\begin{figure}
	\centering
	\includegraphics[width=1.0\textwidth]{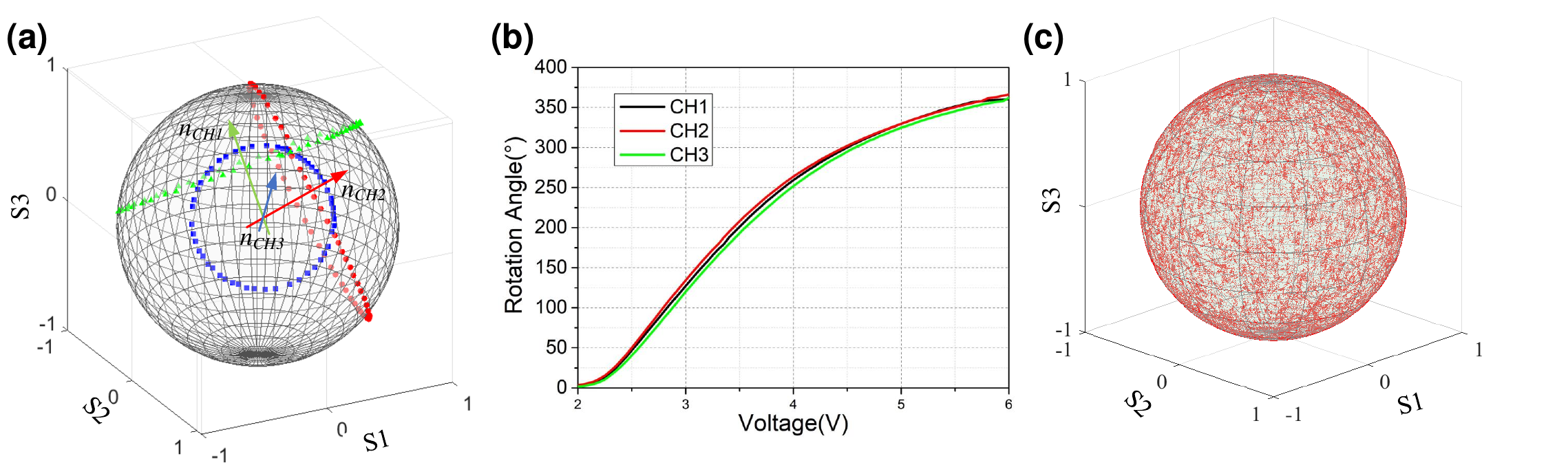}
	\caption{Experimental Results: (a) The polarization modulation characteristics of the three channels; (b) The relationship between the rotation angle and the applied voltage; (c) Poincaré sphere coverage test.}
	\label{FIG:3}
\end{figure}

\begin{enumerate}[1)]
    \item The SOP trajectories generated by the three channels form circular paths on the Poincaré sphere, as shown in Fig.~\ref{FIG:3}(a).
    \item The measured angles between the trajectory normal vectors of CH1, CH2, and CH3 are $92.3^\circ$, $89.0^\circ$, and $91.8^\circ$, respectively, confirming that the three modulation axes are close to the designed orthogonal configuration. The small deviations from the ideal orthogonal relationship are mainly attributed to assembly and alignment tolerances of the liquid-crystal cells.
    \item The trajectory circles revolve around their respective centers. Taking the current SOP as the starting point, the rotation angles corresponding to different voltages can be calculated. The relationship between the rotation angle and the applied voltage is shown in Fig.~\ref{FIG:3}(b).
    \item To evaluate the accessible SOP range of the LC-PC, random voltage combinations were applied to the three channels and the resulting SOPs were mapped onto the Poincaré sphere. Using fine-grid partitioning and density analysis, complete Poincaré-sphere coverage was experimentally verified, demonstrating blind-zone-free polarization control capability, as shown in Fig.~\ref{FIG:3}(c).
\end{enumerate}

\begin{figure}
	\centering
	\includegraphics[width=1.0\textwidth]{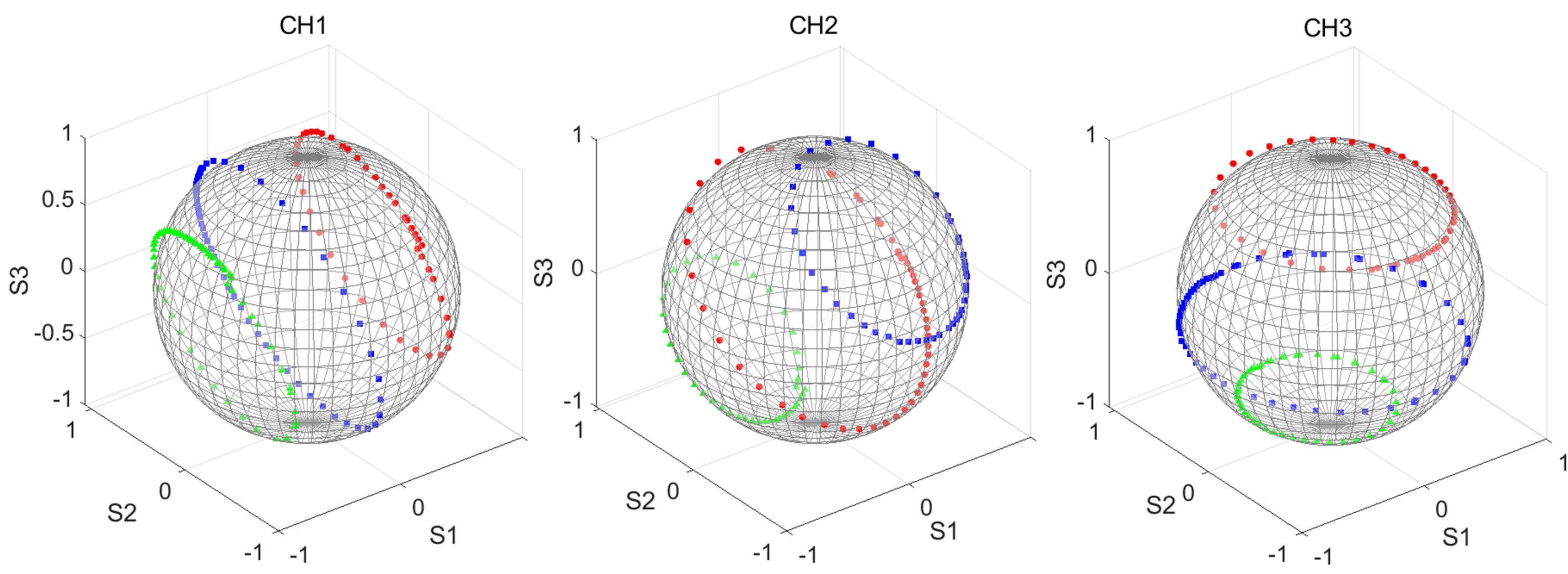}
	\caption{Polarization modulation trajectories on the Poincaré sphere for different input SOPs.}
	\label{FIG:4}
\end{figure}

To further verify the SOP independence of the trajectory normal vectors, the polarization modulation characteristics of each channel were measured under multiple input SOPs. As shown in Fig.~\ref{FIG:4}, different input SOPs produce distinct trajectory circles on the Poincaré sphere. However, the corresponding normal vectors remain highly consistent, with a maximum measured angular difference of $1.806^\circ$ (Table~\ref{tab:table1}), confirming that the trajectory normal vector is an intrinsic property of the liquid-crystal cell and is essentially independent of the input SOP.

\begin{table}
\centering
\caption{Normal vectors of the trajectory circles for different input SOPs.}
\label{tab:table1}
\begin{tabular}{c c ccc c} 
\toprule
Channel & Exp. No. & \multicolumn{3}{c}{Normal vector} & Angle \\
\midrule
\multirow{3}{*}{CH1} & 1 & 1.730 & -0.204 & 0.525 & / \\
                     & 2 & 2.031 & -0.193 & 0.581 & 1.531 \\
                     & 3 & 0.805 & -0.092 & 0.226 & 1.243 \\
\midrule
\multirow{3}{*}{CH2} & 1 & 0.652 & -1.408 & 1.467 & / \\
                     & 2 & 0.371 & -0.784 & 0.867 & 1.669 \\
                     & 3 & 0.182 & -0.371 & 0.413 & 1.806 \\
\midrule
\multirow{3}{*}{CH3} & 1 & 0.738 & 0.780  & 0.555 & / \\
                     & 2 & 1.104 & 1.208  & 0.847 & 0.892 \\
                     & 3 & 0.337 & 0.350  & 0.244 & 0.827 \\
\bottomrule
\end{tabular}

\parbox{\textwidth}{
\raggedright
\footnotesize
Note: Angles were calculated using the original unrounded data, while the values shown in the table are rounded to three decimal places for presentation.
}

\end{table}

The repeatability and hysteresis of the LC-PC were also evaluated. For ten repeated applications of the voltage corresponding to approximately $2\pi$ retardation, the angular deviation relative to the first measurement ranged from $0.114^\circ$ to $0.603^\circ$, with a mean of $0.374^\circ$ and a standard deviation of $0.145^\circ$. Hysteresis was characterized at target voltages of 2, 3, and 4 V by comparing the output SOPs obtained before and after an excursion to 6 V. The corresponding SOP differences were $0.266^\circ$, $0.0747^\circ$, and $0.177^\circ$, respectively, indicating sub-degree cycle-to-cycle variation and hysteresis-induced SOP differences under the tested conditions.

The LC cells are high-resistance capacitive loads and therefore dissipate very little power. The power consumption associated with the LC-PC is dominated by the TEC, which consumes approximately $229\text{ mW}$ in steady state and approximately $405\text{ mW}$ during initial start-up and temperature stabilization under an ambient temperature of $20^\circ\text{C}$ and an operating temperature of $40^\circ\text{C}$.

Because LC birefringence is wavelength dependent, the voltage-to-rotation relationship was further characterized over 1525--1575 nm. This range corresponds to the effective operating band of the antireflection coating used in the present device. The wavelength dependence becomes more pronounced at higher driving voltages. Near the voltage corresponding to approximately $2\pi$ rotation, the rotation-angle difference between 1525 and 1575 nm is approximately $11^\circ$ at the same applied voltage. Therefore, the voltage-to-rotation relationship should be re-characterized when the operating wavelength is changed to maintain degree-level polarization-control accuracy.

\section{Single-measurement direct-calculation polarization control}
\subsection{Principle}
Benefiting from the deterministic polarization response of the fabricated LC-PC, a single-measurement direct-calculation polarization control strategy can be established. As demonstrated in Figs.~\ref{FIG:2} and \ref{FIG:4}, each LC channel exhibits a SOP-independent circular trajectory on the Poincaré sphere with a fixed rotation axis. This property indicates that the LC-PC is not a black-box polarization transformer, but rather a cascade of three deterministic rotations with fixed and well-defined axes.

\begin{figure}
\centering
\includegraphics[width=0.55\linewidth]{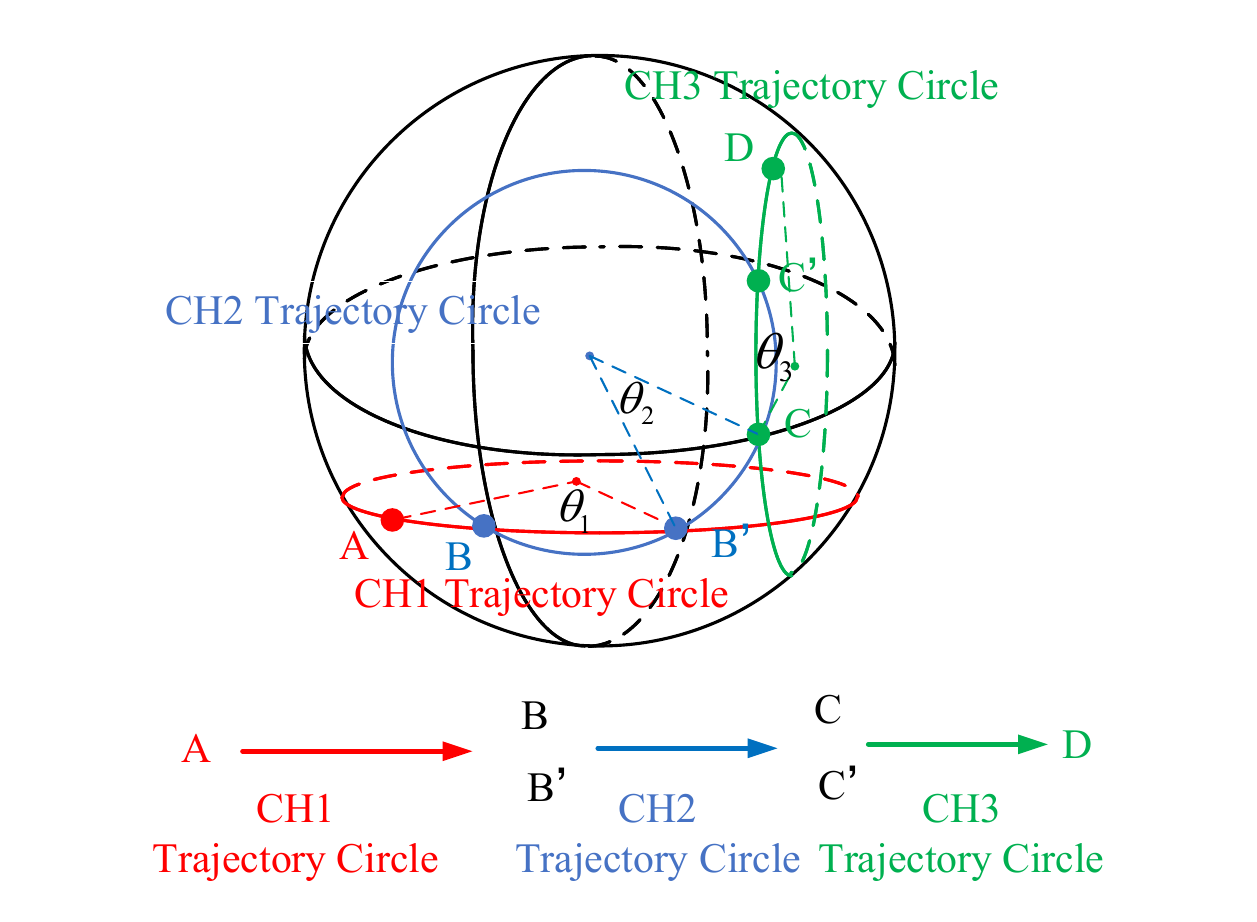}
\caption{Schematic of the polarization control principle.}
\label{FIG:5}
\end{figure}

Based on this physical insight, the polarization control problem is reformulated as a geometric rotation synthesis problem on the Poincaré sphere. As illustrated in Fig.~\ref{FIG:5}, let the initial and target SOPs be denoted by points $A$ and $D$ on the Poincaré sphere, respectively. The transformation from $A$ to $D$ is decomposed into three successive rotations corresponding to the LC channels CH1, CH2, and CH3. 

The initial state $A$ and the CH1 rotation axis $\mathbf{n}_1$ define a circular trajectory constraint on the Poincaré sphere, while the target state $D$ and the CH3 axis $\mathbf{n}_3$ define another one. The CH2 axis $\mathbf{n}_2$ provides the intermediate rotational constraint that couples these two fixed-axis evolutions into a continuous trajectory. By simultaneously solving the corresponding geometric constraints, two sets of intermediate intersection points, $B$ (or $B'$) and $C$ (or $C'$), can be obtained. This establishes a feasible polarization evolution path
\[
{A} \rightarrow {B} \rightarrow C \rightarrow {D} \quad \text{(or} \quad {A} \rightarrow {B}' \rightarrow {C}' \rightarrow {D}\text{)}.
\]

Building on the above formulation, the specific solution procedure is described as follows.

Based on the orthogonality relationships between the rotation axes $\mathbf{n}_1$, $\mathbf{n}_2$, $\mathbf{n}_3$ and the vectors $\mathbf{AB}$, $\mathbf{BC}$, and $\mathbf{CD}$, respectively, and given that points $B$ and $C$ lie on the Poincaré sphere, the following relationships can be established.

\begin{equation}
\begin{cases}
\mathbf{n}_1 \cdot (\mathbf{B} - \mathbf{A}) = 0 \\
\mathbf{n}_2 \cdot (\mathbf{C} - \mathbf{B}) = 0 \\
\mathbf{n}_3 \cdot (\mathbf{D} - \mathbf{C}) = 0 \\
\|\mathbf{B}\| = 1 \\
\|\mathbf{C}\| = 1
\end{cases}
\end{equation}

Equation (3) does not uniquely determine the intermediate SOPs $B$ and $C$, but leaves one degree of freedom and therefore admits multiple feasible polarization-evolution paths. In the implemented algorithm, this degree of freedom is parameterized by discretizing one Stokes component of $B$. The $S_1$ component is normally scanned from $-1$ to $1$ with a step size of 0.01. For each prescribed value, the corresponding solutions of $B$ and $C$ are obtained from Eq. (3), and infeasible solutions are discarded. If the CH1 rotation axis is parallel to the $S_1$ axis, so that $S_1$ remains constant during the CH1 rotation, the $S_2$ component is discretized instead. This numerical discretization is confined to the internal computation and does not require any additional SOP measurements or feedback-based adjustment during a polarization-control operation. 

Once the intermediate SOPs are determined, the required rotation angles of the three LC channels can be calculated from the angular displacement between the corresponding points along their respective circular trajectories on the Poincaré sphere. Specifically, for each LC channel, the rotation angle is obtained by projecting the initial and final SOPs onto the corresponding trajectory plane and calculating their angular separation with respect to the trajectory center. The obtained rotation angles are subsequently converted into driving voltages according to the experimentally characterized voltage-to-rotation relationships of the three LC channels.

Owing to the periodicity of the polarization rotation, the required rotation angle of each channel can be represented within one calibrated $2\pi$ interval and directly mapped to a valid driving voltage. Therefore, successive point-to-point control updates do not accumulate toward the actuator boundaries and can be performed without a separate voltage-reset or state-reinitialization step.

This feedforward formulation enables direct polarization transformation without iterative optimization, gradient-based searching, or feedback-based convergence. The deterministic response of the LC-PC ensures that the calculated rotation angles can be reliably mapped to the required driving voltages. Importantly, once the LC-channel rotation axes and voltage-to-rotation relationships have been characterized, only a single SOP measurement is required to compute the full set of driving voltages, eliminating the need for repeated online SOP measurements, feedback-based iterative convergence, or polarization-transformation-matrix reconstruction during polarization compensation.

\subsection{Experiments and results}
\begin{figure}
	\centering
	\includegraphics[width=0.75\textwidth]{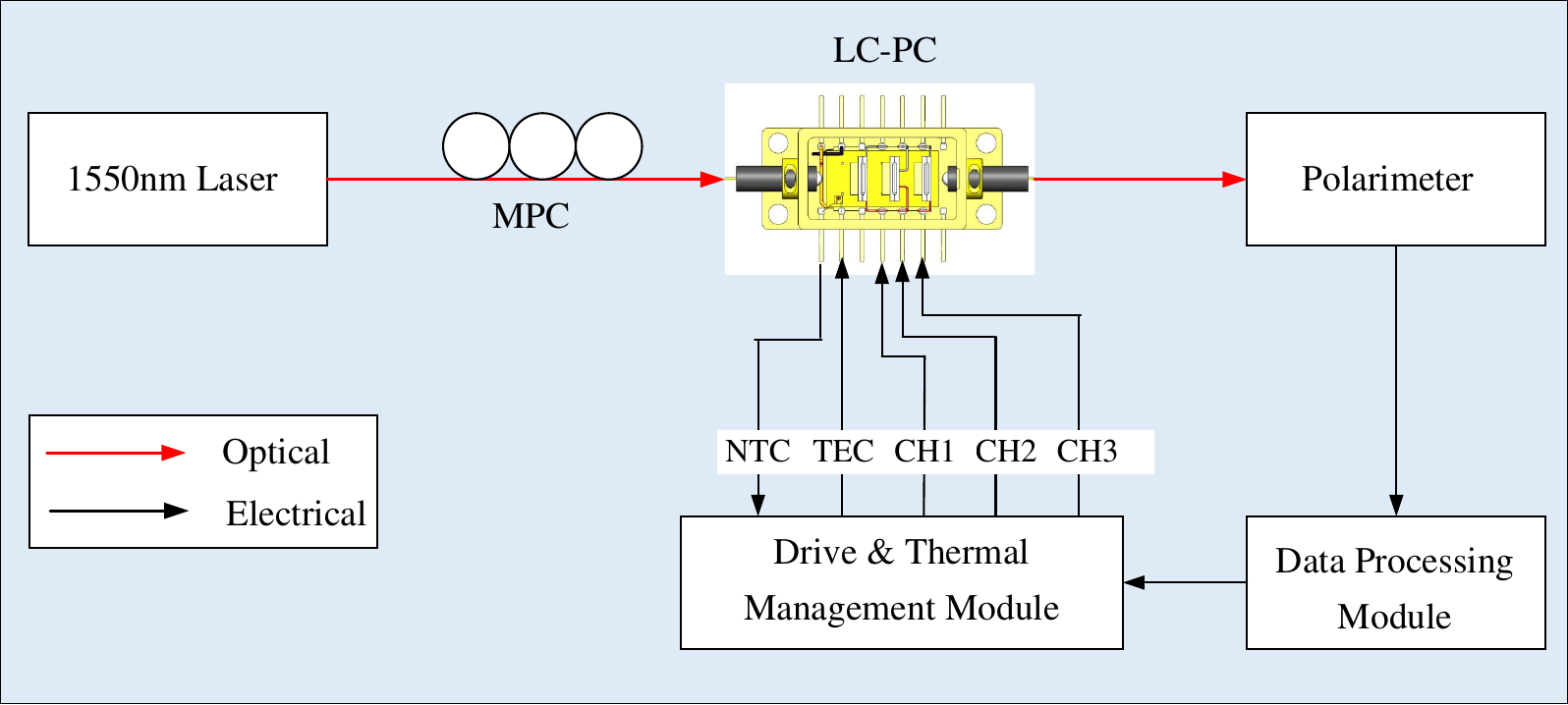}
	\caption{Experimental setup for the verification of the proposed polarization control algorithm.}
	\label{FIG:6}
\end{figure}

To experimentally demonstrate the single-measurement direct-calculation polarization control enabled by the deterministic LC-PC, polarization control experiments were performed using the fabricated device. As shown in Fig.~\ref{FIG:6}, the experimental setup consisted of a 1550-nm laser source, a manual polarization controller (MPC), the LC-PC, a polarimeter, a data processing unit, and a voltage driving module. The laser output was launched sequentially through the MPC and the LC-PC before being detected by the polarimeter. The data processing unit calculated the required driving voltages from the measured Stokes parameters according to the proposed algorithm and sent them to the voltage driver for LC-PC actuation. A Keysight N7788B polarization analyzer was used for SOP measurement; the manufacturer specifies a typical SOP measurement uncertainty of $1.5^\circ$ under the stated operating conditions.

In the experiment, the MPC was used to generate different initial SOPs, and the LC-PC transformed each initial SOP toward a predefined target state. To evaluate the polarization-control performance over different regions of the Poincaré sphere, target SOPs were selected from all eight octants defined by the signs of $S_1$, $S_2$, and $S_3$. For each target SOP, two different initial SOPs were generated using the MPC, resulting in 16 input-to-target transformations. For each case, only a single measurement of the initial SOP was required, after which the driving voltages were directly calculated using the pre-characterized LC-channel parameters and applied without further optimization. No iterative optimization, feedback-based searching, or Mueller-matrix calibration was involved throughout the entire control process.

The resulting SOPs were measured and compared with their corresponding target states. The control accuracy was evaluated by the angular separation between the measured and target SOPs on the Poincaré sphere, calculated using the normalized dot product of the directly measured Stokes vectors. The experimental results are summarized in Table~\ref{tab:table2}. For the 16 tested input-to-target transformations, the measured angular errors ranged from $0.405^\circ$ to $1.966^\circ$.  To further evaluate the repeatability of the polarization-control process, three additional input-to-target SOP combinations were randomly selected and each independently repeated three times under the same experimental conditions. The corresponding mean angular errors were $0.389^\circ$, $0.563^\circ$, and $0.448^\circ$, with standard deviations of $0.272^\circ$, $0.353^\circ$, and $0.057^\circ$, respectively, and all nine target-referenced angular errors were below $1^\circ$. These results demonstrate that the proposed direct-calculation method can accurately transform different input SOPs to prescribed target SOPs distributed across all eight octants of the Poincaré sphere, with degree-level polarization-control accuracy.

\begin{table*}[t]
\centering
\caption{Polarization control results for different initial and target SOPs.}
\label{tab:table2}
\footnotesize
\renewcommand{\arraystretch}{1.08}
\setlength{\tabcolsep}{0pt}

\begin{tabular}{
c
@{\hspace{26pt}}
c
@{\hspace{26pt}}
c
@{\hspace{12pt}}
c
}
\toprule
\multicolumn{3}{c}{\textbf{State of polarization}} &
\multirow{2}{*}{\textbf{Angular error ($^\circ$)}} \\
\cmidrule(lr){1-3}
\textbf{Initial} &
\textbf{Target} &
\textbf{Measured output} &
\\
\midrule

$(-0.339,\,-0.920,\,0.118)$ &
$(0.924,\,0.370,\,0.246)$ &
$(0.922,\,0.374,\,0.266)$ &
1.113 \\

$(-0.016,\,-0.788,\,-0.585)$ &
$(0.924,\,0.370,\,0.246)$ &
$(0.914,\,0.390,\,0.270)$ &
1.812 \\

$(-0.010,\,-0.776,\,-0.601)$ &
$(0.751,\,0.454,\,-0.519)$ &
$(0.736,\,0.466,\,-0.527)$ &
1.167 \\

$(-0.027,\,-0.801,\,-0.571)$ &
$(0.751,\,0.454,\,-0.519)$ &
$(0.733,\,0.470,\,-0.532)$ &
1.537 \\

$(-0.213,\,-0.940,\,-0.206)$ &
$(0.693,\,-0.603,\,0.452)$ &
$(0.675,\,-0.630,\,0.442)$ &
1.901 \\

$(0.282,\,-0.749,\,0.586)$ &
$(0.693,\,-0.603,\,0.452)$ &
$(0.677,\,-0.624,\,0.444)$ &
1.542 \\

$(-0.737,\,-0.422,\,0.512)$ &
$(0.286,\,-0.469,\,-0.813)$ &
$(0.290,\,-0.471,\,-0.807)$ &
0.405 \\

$(-0.433,\,-0.796,\,0.397)$ &
$(0.286,\,-0.469,\,-0.813)$ &
$(0.280,\,-0.493,\,-0.802)$ &
1.579 \\

$(0.794,\,0.282,\,0.591)$ &
$(-0.778,\,0.324,\,0.575)$ &
$(-0.769,\,0.312,\,0.594)$ &
1.359 \\

$(0.783,\,0.358,\,0.569)$ &
$(-0.778,\,0.324,\,0.575)$ &
$(-0.782,\,0.331,\,0.565)$ &
0.721 \\

$(-0.779,\,0.141,\,-0.602)$ &
$(-0.439,\,0.454,\,-0.789)$ &
$(-0.433,\,0.458,\,-0.790)$ &
0.413 \\

$(-0.500,\,0.190,\,0.869)$ &
$(-0.439,\,0.454,\,-0.789)$ &
$(-0.429,\,0.456,\,-0.794)$ &
0.643 \\

$(0.288,\,-0.419,\,0.865)$ &
$(-0.367,\,-0.583,\,0.757)$ &
$(-0.360,\,-0.597,\,0.752)$ &
0.913 \\

$(0.159,\,-0.546,\,0.822)$ &
$(-0.367,\,-0.583,\,0.757)$ &
$(-0.385,\,-0.552,\,0.760)$ &
1.966 \\

$(-0.746,\, -0.581,\,0.288)$ &
$(-0.576,\,-0.651,\,-0.454)$ &
$(-0.595,\,-0.635,\,-0.471)$ &
1.674 \\

$(0.488,\,0.805,\,0.497)$ &
$(-0.576,\,-0.651,\,-0.454)$ &
$(-0.562,\,-0.649,\,-0.433)$ &
0.978 \\

\bottomrule
\end{tabular}
\end{table*}

To further analyze the polarization-control error, the short-term repeatability of the SOP measurement was evaluated by repeatedly measuring a fixed SOP 20 times under unchanged optical and driving conditions. The mean angular deviation was $0.166^\circ$, with a standard deviation of $0.093^\circ$. This measurement variation is relatively small compared with the overall control error, indicating that the latter is also influenced by device-related non-idealities. Assembly and alignment tolerances may cause the measured SOP trajectories to deviate from ideal circular trajectories, thereby affecting the extracted rotation axes and calculated rotation angles. Additional contributions may arise from the voltage-to-rotation characterization and residual SOP measurement variations.

Similarly, using wavelength-specific voltage-to-rotation calibrations, polarization-control experiments at 1525, 1540, 1560, and 1575 nm also achieved angular errors below $2^\circ$ for all tested cases, confirming that the proposed control method remains applicable over the characterized wavelength range after wavelength-specific calibration.

\section{Comparison and discussion}

Table~\ref{tab:table3} compares representative PCs in terms of both device performance and control capability. In addition to conventional metrics, including insertion loss, driving voltage, and footprint, the comparison also considers the associated control strategy and SOP measurement requirements. Such an evaluation provides a more comprehensive understanding of practical polarization controllers, since their performance depends on both the hardware platform and the control algorithm.

From the device perspective, the proposed LC-PC achieves a favorable balance among optical performance, electrical driving requirements, and compactness. It provides a fiber-to-fiber insertion loss of approximately 0.97\,dB, a driving voltage below 6\,V, and a compact butterfly package of $30 \times 12.7 \times 13$\,mm$^3$, while achieving deterministic and repeatable polarization modulation over the full Poincaré sphere. Compared with waveplate- and fiber-squeezer-based PCs \cite{shao-kaiRealizationArbitraryInverse2007,huAdaptivePolarizationControl2023}, the proposed device eliminates mechanical moving components and provides electronically controlled polarization modulation with improved compactness and stability. Compared with integrated PCs based on thin-film lithium niobate (TFLN) \cite{linHighperformancePolarizationManagement2022a}, thin-film lithium tantalate (TFLT)\cite{gaoUltrafastResetfreeIntegrated2026} and silicon photonics (SOI) \cite{qian-ruLowlossIntegratedDynamic2024,zhaoUltracompactSiliconOnchip2024a}, these platforms can provide high-speed polarization-control capability and compact integration. In contrast, the proposed LC-PC avoids additional fiber-to-chip coupling loss by maintaining a fiber-to-fiber optical path, which is advantageous for insertion-loss-sensitive applications such as quantum key distribution systems (QKD). 

From the control perspective, direct-calculation approaches can achieve higher efficiency than gradient-based optimization methods by directly solving the required polarization transformation. However, existing direct-calculation methods generally require multiple SOP measurements to characterize the polarization transformation induced by the transmission channel and reconstruct the corresponding Mueller matrix, based on which the required compensation parameters can be calculated \cite{jimenez-girelaPolarizationcorrectionDeviceLiquid2025a}. In this work, the deterministic polarization response of the LC-PC is exploited to establish a geometric rotation model. The SOP-independent trajectory normal vectors of the LC channels provide fixed rotational axes on the Poincaré sphere, enabling an analytical mapping between polarization states and phase retardations. Therefore, using the pre-characterized LC-channel parameters, the required driving voltages can be directly calculated from a single measured input SOP, avoiding repeated online SOP measurements, feedback-based iterative convergence, and polarization-transformation-matrix reconstruction. In addition, the periodic polarization response allows successive point-to-point control updates to be performed without a separate voltage-reset or state-reinitialization step, avoiding the boundary accumulation that may occur in incremental iterative control.

\begin{table*}[t]
\caption{Comparison of representative polarization controllers in terms of device performance and control capability.}
\label{tab:table3}
\centering
\footnotesize
\renewcommand{\arraystretch}{1.2}

\begin{tabular}{llccccc}
\hline

\textbf{Technology} &
\textbf{Product/} &
\textbf{Insertion} &
\textbf{Driving} &
\textbf{Footprint} &
\textbf{Direct} &
\textbf{Required SOP} \\

\textbf{Platform} &
\textbf{Reference} &
\textbf{Loss (dB)} &
\textbf{Voltage} &
\textbf{(mm)} &
\textbf{Calculation} &
\textbf{Measurements} \\

\hline

Waveplate &
Ref.~\cite{shao-kaiRealizationArbitraryInverse2007} &
\begin{tabular}[c]{@{}c@{}}
$\sim$0.3\\
(Waveplate-only)
\end{tabular} &
N/A &
Discrete optics &
Yes &
Multiple \\

Fiber Squeezer &
Ref.~\cite{huAdaptivePolarizationControl2023} &
0.05 &
$\sim$70 V &
$65.5 \times 20.3 \times 16$ &
No &
N/A \\

TFLN &
Ref.~\cite{linHighperformancePolarizationManagement2022a} &
2.62 &
$\sim$4.8 V &
$15 \times 3^{\mathrm{a}}$ &
No &
N/A \\

TFLT &
Ref.~\cite{gaoUltrafastResetfreeIntegrated2026} &
\begin{tabular}[c]{@{}c@{}}
$<4$\\
(Chip-only)
\end{tabular} &
$V_{\pi}=2.46$ V &
$14 \times 1.3^{\mathrm{a}}$ &
No &
N/A \\

SOI &
Ref.~\cite{qian-ruLowlossIntegratedDynamic2024} &
5.7 &
$\sim$10 V &
$5.2 \times 0.12 \times 0.8$ &
No &
N/A \\

SOI &
Ref.~\cite{zhaoUltracompactSiliconOnchip2024a} &
\begin{tabular}[c]{@{}c@{}}
$<1$\\
(Chip-to-chip)
\end{tabular} &
N/A &
$0.15 \times 0.7^{\mathrm{a}}$ &
No &
N/A \\

Liquid crystal &
Ref.~\cite{jimenez-girelaPolarizationcorrectionDeviceLiquid2025a} &
\begin{tabular}[c]{@{}c@{}}
$\sim$0.5\\
(Cell-only)
\end{tabular} &
$\sim$4 V &
Discrete optics &
Yes &
Multiple \\

Liquid crystal &
\textbf{This work} &
\textbf{$\sim$0.97} &
\textbf{$<6$ V} &
\textbf{$30 \times 12.7 \times 13$} &
\textbf{Yes} &
\textbf{Single} \\

\hline
\end{tabular}

\vspace{0.2mm}

\begin{flushleft}
\footnotesize
$^{\mathrm{a}}$ Footprint corresponds to the bare (unpackaged) chip; the fully packaged device will have a significantly larger footprint.\\
Unless otherwise specified, insertion loss values correspond to fiber-to-fiber measurements.
\end{flushleft}
\end{table*}

The direct-calculation algorithm eliminates feedback-based iterative convergence; however, the overall control speed is still limited by SOP acquisition, computation, and the physical response of the LC-PC. In the present laboratory implementation, the control algorithm was executed in MATLAB on a standard PC. The SOP acquisition, algorithm computation, and voltage-driver response required approximately $0.4$ ms, $105$ ms, and $1.5~\mu$s, respectively, while the LC response time ranged from approximately $102$ to $946$ ms at $40^\circ$C depending on the voltage transition. The resulting point-to-point control-update time was therefore approximately $207$--$1051$ ms, corresponding to a nominal update rate of approximately $0.95$--$4.8$ Hz under the tested conditions. It should be noted that the reported computation time reflects the current implementation on a general-purpose PC; the algorithm itself can be readily ported to an embedded processor or FPGA platform to substantially reduce the computational latency.

The proposed device and control algorithm are highly synergistic. The compact deterministic LC-PC provides a robust and practical platform, while the single-measurement control scheme significantly reduces deployment complexity. This combination is especially attractive for loss- and space-constrained polarization-sensitive photonic systems, such as QKD terminals, optical sensing, and biomedical imaging. The primary limitation of the device remains the LC response speed, which depends on the applied voltage transition and operating temperature. Future work will focus on adopting fast-response LC materials (e.g., ferroelectric LCs)\cite{guoReverseBistableEffect2015} and optimized cell/packaging designs to further improve response speed and compactness.

\section{Conclusion}
A compact butterfly-packaged deterministic LC-PC has been demonstrated together with a single-measurement direct-calculation polarization control algorithm. The fabricated LC-PC achieves low insertion loss, low driving voltage, and full Poincaré sphere coverage while exhibiting deterministic polarization modulation with SOP-independent trajectory normal vectors. Leveraging these intrinsic characteristics, the proposed algorithm establishes a direct geometric relationship between the measured SOP and the required control signals, enabling degree-level polarization-control accuracy using pre-characterized device parameters and a single SOP measurement, without iterative optimization.

The combination of the compact deterministic LC-PC and the single-measurement direct-calculation algorithm provides an efficient solution for size-, weight-, and loss-constrained polarization-sensitive photonic systems, including QKD terminals, optical sensing, and photonic systems. Future work will focus on improving response speed through faster liquid-crystal materials and optimized packaging, followed by system-level reliability evaluation, fiber-based QKD demonstrations, and radiation-hardness verification toward spaceborne applications.

\printcredits

\bibliographystyle{model1-num-names}

\bibliography{cas-refs}

@article{aartsNewEndlessPolarization1989,
  title = {New Endless Polarization Control Method Using Three Fiber Squeezers},
  author = {Aarts, W.H.J. and Khoe, G.-D.},
  year = 1989,
  month = jul,
  journal = {Journal of Lightwave Technology},
  volume = {7},
  number = {7},
  pages = {1033--1043},
  issn = {1558-2213},
  doi = {10.1109/50.29630},
  urldate = {2026-06-11}
}

@article{chenEnhancingWeakDisturbance2025,
  title = {Enhancing Weak Disturbance Localization in Asymmetric Optical Fiber Vibration Sensing System Based on Polarization Control},
  author = {Chen, Chaoxiang and Liu, Kun and Jing, Jianying and Hu, Xinxin and Xue, Kang and Zhang, Dongqi and Du, Ziwen and Li, Guixian and Xu, Tianhua and Jiang, Junfeng and Liu, Tiegen},
  year = 2025,
  month = dec,
  journal = {Optics and Lasers in Engineering},
  volume = {195},
  pages = {109249},
  issn = {0143-8166},
  doi = {10.1016/j.optlaseng.2025.109249},
  urldate = {2026-06-11}
}

@article{goldsteinAnalysisPolarizedLight1970a,
  title = {Analysis of Polarized Light with Two Quarter-Wave Plates},
  author = {Goldstein, D. J.},
  year = 1970,
  journal = {Journal of Microscopy},
  volume = {91},
  number = {1},
  pages = {19--30},
  issn = {1365-2818},
  doi = {10.1111/j.1365-2818.1970.tb02200.x},
  urldate = {2026-06-13},
  copyright = {1970 Blackwell Science Ltd},
  langid = {english}
}

@article{guanReviewPolarizationbasedTechnology2025,
  title = {Review of Polarization-Based Technology for Biomedical Applications},
  author = {Guan, Caizhong and Zeng, Nan and He, Honghui},
  year = 2025,
  month = mar,
  journal = {Journal of Innovative Optical Health Sciences},
  volume = {18},
  number = {02},
  pages = {2430002},
  publisher = {World Scientific Publishing Co.},
  issn = {1793-5458},
  doi = {10.1142/S1793545824300027},
  urldate = {2026-06-13}
}

@article{guoReverseBistableEffect2015,
  title = {Reverse Bistable Effect in Ferroelectric Liquid Crystal Devices with Ultra-Fast Switching at Low Driving Voltage},
  author = {Guo, Qi and Zhao, Xiaojin and Zhao, Huijie and Chigrinov, V. G.},
  year = 2015,
  month = may,
  journal = {Optics Letters},
  volume = {40},
  number = {10},
  pages = {2413--2416},
  publisher = {Optica Publishing Group},
  issn = {1539-4794},
  doi = {10.1364/OL.40.002413},
  urldate = {2026-06-11},
  copyright = {\copyright{} 2015 Optical Society of America},
  langid = {english}
}

@article{hanProgrammableLinearRetarder2025,
  title = {Programmable Linear Retarder Using Liquid Crystal Variable Retarders with Calibration for Non-Ideal Beam Splitter Properties},
  author = {Han, Chien-Yuan and Chao, Zhen-Xiang and Chan, Yi-Hsin and Yu, Chih-Jen},
  year = 2025,
  month = jan,
  journal = {Optics and Lasers in Engineering},
  volume = {184},
  pages = {108623},
  issn = {0143-8166},
  doi = {10.1016/j.optlaseng.2024.108623},
  urldate = {2026-06-11}
}

@article{hePolarisationOpticsBiomedical2021a,
  title = {Polarisation Optics for Biomedical and Clinical Applications: A Review},
  shorttitle = {Polarisation Optics for Biomedical and Clinical Applications},
  author = {He, Chao and He, Honghui and Chang, Jintao and Chen, Binguo and Ma, Hui and Booth, Martin J.},
  year = 2021,
  month = sep,
  journal = {Light: Science \& Applications},
  volume = {10},
  number = {1},
  pages = {194},
  publisher = {Nature Publishing Group},
  issn = {2047-7538},
  doi = {10.1038/s41377-021-00639-x},
  urldate = {2026-06-11},
  copyright = {2021 The Author(s)},
  langid = {english}
}

@article{hirabayashiFeedforwardContinuousComplete2003,
  title = {Feed-Forward Continuous and Complete Polarization Control with a {{PLZT}} Rotatable-Variable Waveplate and Inline Polarimeter},
  author = {Hirabayashi, K. and Amano, C.},
  year = 2003,
  month = sep,
  journal = {Journal of Lightwave Technology},
  volume = {21},
  number = {9},
  pages = {1920--1932},
  issn = {1558-2213},
  doi = {10.1109/JLT.2003.816891},
  urldate = {2026-06-11}
}

@article{jimenez-girelaPolarizationcorrectionDeviceLiquid2025a,
  title = {Polarization-Correction Device with Liquid Crystals for Quantum-Key-Distribution Satellite Systems},
  author = {{Jimenez-Girela}, A. and {Merino-P{\'e}rez}, D. and {Campos-Jara}, A. and Negr{\'i}n, J. Socas and Parejo, P. Garcia and {\'A}lvarez-Herrero, A.},
  year = 2025,
  month = jun,
  journal = {Physical Review Applied},
  volume = {23},
  number = {6},
  pages = {064070},
  publisher = {American Physical Society},
  doi = {10.1103/8plr-m6n8},
  urldate = {2026-06-11}
}

@article{kimIntegratedopticPolarizationControllers2012,
  title = {Integrated-Optic Polarization Controllers Incorporating Polymer Waveguide Birefringence Modulators},
  author = {Kim, Jun-Whee and Park, Su-Hyun and Chu, Woo-Sung and Oh, Min-Cheol},
  year = 2012,
  month = may,
  journal = {Optics Express},
  volume = {20},
  number = {11},
  pages = {12443--12448},
  publisher = {Optica Publishing Group},
  issn = {1094-4087},
  doi = {10.1364/OE.20.012443},
  urldate = {2026-06-11},
  copyright = {\copyright{} 2012 OSA},
  langid = {english}
}

@article{liHighrateQuantumKey2023a,
  title = {High-Rate Quantum Key Distribution Exceeding 110 {{Mb}} s--1},
  author = {Li, Wei and Zhang, Likang and Tan, Hao and Lu, Yichen and Liao, Sheng-Kai and Huang, Jia and Li, Hao and Wang, Zhen and Mao, Hao-Kun and Yan, Bingze and Li, Qiong and Liu, Yang and Zhang, Qiang and Peng, Cheng-Zhi and You, Lixing and Xu, Feihu and Pan, Jian-Wei},
  year = 2023,
  month = may,
  journal = {Nature Photonics},
  volume = {17},
  number = {5},
  pages = {416--421},
  issn = {1749-4885, 1749-4893},
  doi = {10.1038/s41566-023-01166-4},
  urldate = {2025-11-05},
  langid = {english}
}

@article{linHighperformancePolarizationManagement2022a,
  title = {High-Performance Polarization Management Devices Based on Thin-Film Lithium Niobate},
  author = {Lin, Zhongjin and Lin, Yanmei and Li, Hao and Xu, Mengyue and He, Mingbo and Ke, Wei and Tan, Heyun and Han, Ya and Li, Zhaohui and Wang, Dawei and Yao, X. Steve and Fu, Songnian and Yu, Siyuan and Cai, Xinlun},
  year = 2022,
  month = apr,
  journal = {Light: Science \& Applications},
  volume = {11},
  number = {1},
  pages = {93},
  publisher = {Nature Publishing Group},
  issn = {2047-7538},
  doi = {10.1038/s41377-022-00779-8},
  urldate = {2026-06-11},
  copyright = {2022 The Author(s)},
  langid = {english}
}

@article{liuContinuousvariableQuantumKey2020a,
  title = {Continuous-Variable Quantum Key Distribution under Strong Channel Polarization Disturbance},
  author = {Liu, Wenyuan and Cao, Yanxia and Wang, Xuyang and Li, Yongmin},
  year = 2020,
  month = sep,
  journal = {Physical Review A},
  volume = {102},
  number = {3},
  pages = {032625},
  publisher = {American Physical Society},
  doi = {10.1103/PhysRevA.102.032625},
  urldate = {2026-06-13}
}

@article{luoResearchPolarizationCompensation2024,
  title = {Research on Polarization Compensation for Practical Satellite-Based Quantum Key Distribution},
  author = {Luo, Wen-Bin and Li, Yang and Li, Yu-Huai and Tao, Xue-Ying and Chen, Hao-Ze and Hua, An and Cai, Wen-Qi and Yin, Juan and Ren, Ji-Gang and Liao, Sheng-Kai and Peng, Cheng-Zhi},
  year = 2024,
  month = nov,
  journal = {Optics Communications},
  volume = {570},
  pages = {130925},
  issn = {0030-4018},
  doi = {10.1016/j.optcom.2024.130925},
  urldate = {2026-06-11}
}

@article{maAutomatedControlAlgorithms2020,
  title = {Automated Control Algorithms for Silicon Photonic Polarization Receiver},
  author = {Ma, Minglei and Shoman, Hossam and Tang, Keyi and Shekhar, Sudip and Jaeger, Nicolas A. F. and Chrostowski, Lukas},
  year = 2020,
  month = jan,
  journal = {Optics Express},
  volume = {28},
  number = {2},
  pages = {1885--1896},
  publisher = {Optica Publishing Group},
  issn = {1094-4087},
  doi = {10.1364/OE.380121},
  urldate = {2026-06-13},
  copyright = {\copyright{} 2020 Optical Society of America},
  langid = {english}
}

@article{peranicStudyPolarizationCompensation2023,
  title = {A Study of Polarization Compensation for Quantum Networks},
  author = {Perani{\'c}, Matej and Clark, Marcus and Wang, Rui and Bahrani, Sima and Alia, Obada and Wengerowsky, S{\"o}ren and Radman, Anton and Lon{\v c}ari{\'c}, Martin and Stip{\v c}evi{\'c}, Mario and Rarity, John and Nejabati, Reza and Joshi, Siddarth Koduru},
  year = 2023,
  month = aug,
  journal = {EPJ Quantum Technology},
  volume = {10},
  number = {1},
  pages = {30},
  issn = {2196-0763},
  doi = {10.1140/epjqt/s40507-023-00187-w},
  urldate = {2026-06-11},
  langid = {english}
}

@article{qian-ruLowlossIntegratedDynamic2024,
  title = {{Low-loss integrated dynamic polarization controller based on silicon photonics}},
  author = {{Qian-Ru}, Zhao and {Xu-Yang}, Wang and {Yan-Xiang}, Jia and {Yun-Jie}, Zhang and {Zhen-Guo}, Lu and Yi, Qian and Jun, Zou and {Yong-Min}, Li},
  year = 2024,
  month = jan,
  journal = {Acta Physica Sinica},
  volume = {73},
  number = {2},
  pages = {024205--11},
  issn = {1000-3290},
  doi = {10.7498/aps.73.20231214},
  urldate = {2026-06-13},
  copyright = {http://creativecommons.org/licenses/by/3.0/},
  langid = {chinese}
}

@article{sarmiento-merenguelDemonstrationIntegratedPolarization2015,
  title = {Demonstration of Integrated Polarization Control with a 40\,\,{{dB}} Range in Extinction Ratio},
  author = {{Sarmiento-Merenguel}, J. D. and Halir, R. and Roux, X. Le and {Alonso-Ramos}, C. and Vivien, L. and Cheben, P. and {Dur{\'a}n-Valdeiglesias}, E. and {Molina-Fern{\'a}ndez}, I. and {Marris-Morini}, D. and Xu, D.-X. and Schmid, J. H. and Janz, S. and {Ortega-Mo{\~n}ux}, A.},
  year = 2015,
  month = dec,
  journal = {Optica},
  volume = {2},
  number = {12},
  pages = {1019--1023},
  publisher = {Optica Publishing Group},
  issn = {2334-2536},
  doi = {10.1364/OPTICA.2.001019},
  urldate = {2026-06-11},
  copyright = {\copyright{} 2015 Optical Society of America},
  langid = {english}
}

@article{shahwarPolarizationManagementSilicon2024a,
  title = {Polarization Management in Silicon Photonics},
  author = {Shahwar, Dura and Yoon, Hoon Hahn and Akkanen, Suvi-Tuuli and Li, Diao and tul Muntaha, Sidra and Cherchi, Matteo and Aalto, Timo and Sun, Zhipei},
  year = 2024,
  month = sep,
  journal = {npj Nanophotonics},
  volume = {1},
  number = {1},
  pages = {35},
  publisher = {Nature Publishing Group},
  issn = {2948-216X},
  doi = {10.1038/s44310-024-00033-6},
  urldate = {2026-06-11},
  copyright = {2024 The Author(s)},
  langid = {english}
}

@article{tangDistributedOpticalFiber2024,
  title = {Distributed Optical Fiber Magnetic Field Sensor Based on Polarization-Sensitive {{OFDR}}},
  author = {Tang, Yi and Zhu, Mengshi and Pang, Fufei and Wei, Heming and Zhang, Liang and Chen, Wei and Wang, Tingyun},
  year = 2024,
  month = mar,
  journal = {Optics Express},
  volume = {32},
  number = {7},
  pages = {11726--11736},
  publisher = {Optica Publishing Group},
  issn = {1094-4087},
  doi = {10.1364/OE.511557},
  urldate = {2026-06-13},
  copyright = {\copyright{} 2024 Optica Publishing Group},
  langid = {english}
}

@article{tanPolarizationCompensationMethod2023,
  title = {Polarization Compensation Method Based on the Wave Plate Group in Phase Mismatch for Free-Space Quantum Key Distribution},
  author = {Tan, Yongjian and Zhang, Liang and Sun, Tianxing and Song, Zhihua and Wu, Jincai and He, Zhiping},
  year = 2023,
  month = feb,
  journal = {EPJ Quantum Technology},
  volume = {10},
  number = {1},
  pages = {6},
  issn = {2196-0763},
  doi = {10.1140/epjqt/s40507-023-00163-4},
  urldate = {2026-06-13},
  langid = {english}
}

@article{tanRealtimePolarizationCompensation2024a,
  title = {Real-Time Polarization Compensation Method in Quantum Communication Based on Channel {{Muller}} Parameters Detection},
  author = {Tan, Yongjian and Wang, Jianyu and Wu, Jincai and He, Zhiping},
  year = 2024,
  month = mar,
  journal = {Communications Engineering},
  volume = {3},
  number = {1},
  pages = {57},
  publisher = {Nature Publishing Group},
  issn = {2731-3395},
  doi = {10.1038/s44172-024-00198-0},
  urldate = {2026-06-11},
  copyright = {2024 The Author(s)},
  langid = {english}
}

@article{wuResearchRealtimePolarization2022,
  title = {Research on a Real-Time Polarization Compensation Method for Dynamic Quantum Communication Terminals},
  author = {Wu, Jincai and Tan, Yongjian and Zhang, Liang and Dou, Yonghao and Song, Zhihua and Jia, Jianjun and Shu, Rong and He, Zhiping and Wang, Jianyu},
  year = 2022,
  month = feb,
  journal = {Optics and Lasers in Engineering},
  volume = {149},
  pages = {106794},
  issn = {0143-8166},
  doi = {10.1016/j.optlaseng.2021.106794},
  urldate = {2026-06-13}
}

@article{yinEntanglementbasedSecureQuantum2020,
  title = {Entanglement-Based Secure Quantum Cryptography over 1,120 Kilometres},
  author = {Yin, Juan and Li, Yu-Huai and Liao, Sheng-Kai and Yang, Meng and Cao, Yuan and Zhang, Liang and Ren, Ji-Gang and Cai, Wen-Qi and Liu, Wei-Yue and Li, Shuang-Lin and Shu, Rong and Huang, Yong-Mei and Deng, Lei and Li, Li and Zhang, Qiang and Liu, Nai-Le and Chen, Yu-Ao and Lu, Chao-Yang and Wang, Xiang-Bin and Xu, Feihu and Wang, Jian-Yu and Peng, Cheng-Zhi and Ekert, Artur K. and Pan, Jian-Wei},
  year = 2020,
  month = jun,
  journal = {Nature},
  volume = {582},
  number = {7813},
  pages = {501--505},
  publisher = {Nature Publishing Group},
  issn = {1476-4687},
  doi = {10.1038/s41586-020-2401-y},
  urldate = {2026-06-13},
  copyright = {2020 The Author(s), under exclusive licence to Springer Nature Limited},
  langid = {english}
}

@article{yinPolarizationCompensationEntanglementbased2025a,
  title = {Polarization Compensation for Entanglement-Based Quantum Key Distribution},
  author = {Yin, Shi-Hua and Luo, Tian-Wang and Jiang, Wen-Juan and Liu, Xi-Yu and Yu, Ya-Fei and Wei, Zheng-Jun and Zhao, Tian-Ming and Fang, Jun-Bin and Wang, Jin-Dong},
  year = 2025,
  month = sep,
  journal = {Optics Express},
  volume = {33},
  number = {18},
  pages = {38431--38439},
  publisher = {Optica Publishing Group},
  issn = {1094-4087},
  doi = {10.1364/OE.573091},
  urldate = {2026-06-13},
  copyright = {\copyright{} 2025 Optica Publishing Group},
  langid = {english}
}

@article{youngComparisonOptimisationAlgorithms2026,
  title = {A {{Comparison}} of {{Optimisation Algorithms}} for {{Electronic Polarisation Control}} in {{Quantum Key Distribution}}},
  author = {Young, Matt and Duan, Haofan and Pirandola, Stefano and Lucamarini, Marco},
  year = 2026,
  month = jan,
  journal = {Applied Sciences},
  volume = {16},
  number = {5},
  pages = {2568},
  publisher = {Multidisciplinary Digital Publishing Institute},
  issn = {2076-3417},
  doi = {10.3390/app16052568},
  urldate = {2026-06-11},
  copyright = {http://creativecommons.org/licenses/by/3.0/},
  langid = {english}
}

@article{zhangParticleSwarmOptimization2005,
  title = {Particle Swarm Optimization Used as a Control Algorithm for Adaptive {{PMD}} Compensation},
  author = {Zhang, Xiaoguang and Zheng, Yuan and Shen, Yu and Zhang, Jianzhong and Yang, Bojun},
  year = 2005,
  month = jan,
  journal = {IEEE Photonics Technology Letters},
  volume = {17},
  number = {1},
  pages = {85--87},
  issn = {1941-0174},
  doi = {10.1109/LPT.2004.838150},
  urldate = {2026-06-11}
}

@article{zhaoUltracompactSiliconOnchip2024a,
  title = {Ultracompact Silicon On-Chip Polarization Controller},
  author = {Zhao, Weike and Peng, Yingying and Zhu, Mingyu and Liu, Ruoran and Hu, Xiaolong and Shi, Yaocheng and Dai, Daoxin},
  year = 2024,
  month = feb,
  journal = {Photonics Research},
  volume = {12},
  number = {2},
  pages = {183--193},
  publisher = {Optica Publishing Group},
  issn = {2327-9125},
  doi = {10.1364/PRJ.499801},
  urldate = {2026-06-11},
  copyright = {\copyright{} 2024 Chinese Laser Press},
  langid = {english}
}

@article{zhuangPolarizationControllerUsing1999,
  title = {Polarization Controller Using Nematic Liquid Crystals},
  author = {Zhuang, Zhizhong and Suh, Seong-Woo and Patel, J. S.},
  year = 1999,
  month = may,
  journal = {Optics Letters},
  volume = {24},
  number = {10},
  pages = {694--696},
  publisher = {Optica Publishing Group},
  issn = {1539-4794},
  doi = {10.1364/OL.24.000694},
  urldate = {2026-06-11},
  copyright = {\copyright{} 1999 Optical Society of America},
  langid = {english}
}

@article{duSiliconbasedDecoderPolarizationencoding2023,
  title = {Silicon-Based Decoder for Polarization-Encoding Quantum Key Distribution},
  author = {Du, Yongqiang and Zhu, Xun and Hua, Xin and Zhao, Zhengeng and Hu, Xiao and Qian, Yi and Xiao, Xi and Wei, Kejin},
  year = 2023,
  month = mar,
  journal = {Chip},
  volume = {2},
  number = {1},
  pages = {100039},
  issn = {2709-4723},
  doi = {10.1016/j.chip.2023.100039},
  urldate = {2026-06-13}
}

@article{weiResourceefficientQuantumKey2023a,
  title = {Resource-Efficient Quantum Key Distribution with Integrated Silicon Photonics},
  author = {Wei, Kejin and Hu, Xiao and Du, Yongqiang and Hua, Xin and Zhao, Zhengeng and Chen, Ye and Huang, Chunfeng and Xiao, Xi},
  year = 2023,
  month = aug,
  journal = {Photonics Research},
  volume = {11},
  number = {8},
  pages = {1364--1372},
  publisher = {Optica Publishing Group},
  issn = {2327-9125},
  doi = {10.1364/PRJ.482942},
  urldate = {2026-06-13},
  copyright = {\copyright{} 2023 Chinese Laser Press},
  langid = {english}
}

@book{Yeh1999,
  author    = {Pochi Yeh and Claire Gu},
  title     = {Optics of Liquid Crystal Displays},
  publisher = {Wiley},
  year       = {1999}
}

@article{shao-kaiRealizationArbitraryInverse2007,
  title = {Realization of {{Arbitrary Inverse Unitary Transformation}} of {{Single Mode Fibre}} by {{Using Three Wave Plates}}},
  author = {{Shao-Kai}, Wang and {Ji-Gang}, Ren and {Cheng-Zhi}, Peng and Shuo, Jiang and {Xiang-Bin}, Wang},
  year = 2007,
  month = sep,
  journal = {Chinese Physics Letters},
  volume = {24},
  number = {9},
  pages = {2471},
  issn = {0256-307X},
  doi = {10.1088/0256-307X/24/9/003},
  urldate = {2026-07-20},
  langid = {english}
}

@article{huAdaptivePolarizationControl2023,
  title = {Adaptive Polarization Control for a Fiber System Based on the Optimized {{AdamSPGD}} Algorithm},
  author = {Hu, Chen and Luo, Bin and Pan, Wei and Yan, Lianshan and Zou, Xihua},
  year = 2023,
  month = nov,
  journal = {Applied Optics},
  volume = {62},
  number = {33},
  pages = {8798--8803},
  publisher = {Optica Publishing Group},
  issn = {2155-3165},
  doi = {10.1364/AO.503759},
  urldate = {2026-07-20},
  copyright = {\copyright{} 2023 Optica Publishing Group},
  langid = {english}
}

@article{gaoUltrafastResetfreeIntegrated2026,
  title = {Ultrafast Reset-Free Integrated Polarization Tracking Enabled by Thin-Film Lithium Tantalate},
  author = {Gao, Zichao and Lu, Siyu and Zhang, Mingming and Yao, Gengqi and Zhang, Chicheng and Deng, Miao and Chen, Siyu and Dai, Yiqi and Yue, Shiqi and Li, Chijun and Li, Yuqi and Zhou, Ziwen and Liu, Zheli and Yu, Xinyang and Ji, Xitao and Zeng, Cheng and Yan, Siqi and Xia, Jinsong and Tang, Ming},
  year = 2026,
  month = jul,
  journal = {Nature Communications},
  volume = {17},
  number = {1},
  pages = {8761},
  publisher = {Nature Publishing Group},
  issn = {2041-1723},
  doi = {10.1038/s41467-026-75732-z},
  urldate = {2026-09-26},
  copyright = {2026 The Author(s)},
  langid = {english}
}


\end{document}